\documentclass[aps,prd,nofootinbib,amsmath,amssymb,superscriptaddress,showpacs,showkeys,twocolumn,10pt]{revtex4}%superscriptaddress,showpacs,
\usepackage{graphicx}
\usepackage{dcolumn}
\usepackage{bm}
\usepackage{amssymb}
\usepackage{latexsym}
\usepackage{booktabs}
\usepackage{amsmath}
\usepackage{multirow}
\usepackage[colorlinks=true, linkcolor=red, citecolor=blue]{hyperref}
\usepackage{color}
\newcommand{\be}{\begin{equation}}
\newcommand{\ee}{\end{equation}}
\newcommand{\bq}{\begin{eqnarray}}
\newcommand{\eq}{\end{eqnarray}}

\begin{document}

\title{Redshift drift constraints on $f(T)$ gravity}

\author{Jia-Jia Geng}
\affiliation{Department of Physics, College of Sciences, Northeastern University, Shenyang
110004, China}
\author{Rui-Yun Guo}
\affiliation{Department of Physics, College of Sciences, Northeastern University, Shenyang
110004, China}
\author{Dong-Ze He}
\affiliation{Department of Physics, College of Sciences, Northeastern University, Shenyang
110004, China}
\author{Jing-Fei Zhang}
\affiliation{Department of Physics, College of Sciences, Northeastern University, Shenyang
110004, China}
\author{Xin Zhang\footnote{Corresponding author}}
\email{zhangxin@mail.neu.edu.cn} \affiliation{Department of Physics, College of Sciences,
Northeastern University, Shenyang 110004, China}
\affiliation{Center for High Energy Physics, Peking University, Beijing 100080, China}

\begin{abstract}
We explore the impact of the Sandage-Loeb (SL) test on the precision of cosmological constraints for $f(T)$ gravity theories. The SL test is an important supplement to current cosmological observations because it measures the redshift drift in the Lyman-$\alpha$ forest in the spectra of distant quasars, covering the ``redshift desert" of $2 \lesssim z \lesssim5$. To avoid data inconsistency, we use the best-fit models based on current combined observational data as fiducial models to simulate 30 mock SL test data. We quantify the impact of these SL test data on parameter estimation for $f(T)$ gravity theories. Two typical $f(T)$ models are considered, the power-law model $f(T)_{PL}$ and the exponential-form model $f(T)_{EXP}$. The results show that the SL test can effectively break the existing strong degeneracy between the present-day matter density $\Omega_m$ and the Hubble constant $H_0$ in other cosmological observations. For the considered $f(T)$ models, a 30-year observation of the SL test can improve the constraint precision of $\Omega_m$ and $H_0$ enormously but cannot effectively improve the constraint precision of the model parameters. 
\end{abstract}

\pacs{95.36.+x, 98.80.Es, 98.80.-k}
\keywords{redshift drift, cosmological constraints, dark energy, modified gravity, $f(T)$ gravity}

\maketitle

\section{Introduction}

Redshift drift observation directly measures the expansion rate of the universe in the ``redshift desert" of $2 \lesssim z \lesssim5$, which is not covered by existing cosmological observations. Therefore, it could be an important supplement to the other cosmological observations.  In 1962, Sandage first proposed to directly measure the variation of the redshift of distant sources~\cite{sandage}. Then in 1998, Loeb suggested the possibility of detecting redshift drift by decades-long observation of the Lyman-$\alpha$ forest in the spectra of distant quasars (quasi-stellar objects, QSOs) \cite{loeb}. Thus, redshift drift measurement is also referred to as the Sandage-Loeb (SL) test. The 39-m European Extremely Large Telescope (E-ELT) (under construction) is equipped with a high-resolution spectrograph called CODEX (COsmic Dynamics EXperiment), which is designed to collect such SL test signals. A great amount of work has been done on the effect of the SL test on cosmological parameter estimation~\cite{sl1,sl2,sl3,sl4,sl5,sl6,sl7}. As far as we know, in most existing works, the best-fit $\Lambda$ cold dark matter model is usually chosen as the fiducial model in simulating the mock SL test data. When these simulated SL test data are combined with other actual data, tension may exist inside the combined data.

In our previous works~\cite{msl1, msl2, msl3}, we quantified the impact of future redshift drift measurement on parameter estimation for different dark energy models.
This work was based on 30 QSOs, because only about 30 QSOs will be bright enough or lie at a high enough redshift to allow observation of the redshift drift using a telescope such as the E-ELT, according to a Monto Carlo simulation~\cite{Liske}.
To simulate mock SL test data that are consistent with other actual observations, we choose the best-fit dark energy models as the fiducial models in the fit to current data. 
We find that the SL test data alone cannot tightly constrain dark energy models because of the lack of low-redshift data.
However, when combined with other actual observations, the SL test can effectively break the existing parameter degeneracies in current observations and greatly improve the precision of parameter estimation for widely studied dark energy models~\cite{msl1, msl2, msl3}.

It is well known that aside from the theory of dark energy~\cite{quint,phantom,k,Chaplygin,ngcg,HDE1,HDE2,HDE3,HDE4,tachyonic,hessence,YMC,hscalar,PPF1,PPF2,DE1,DE2,DE3,DE4,DE5,DE6,DEW,DEN1,DEN2,limiao},
other explanations for cosmic acceleration exists; one of the most popular is a modification of Einstein's general relativity, i.e.,
modified gravity (MG)~\cite{SH,PR,DGP,GB,Galileon,FR1,FR2,FT1,FT2,FRT,wupx,huangqg,MG1,MG2,MG3,MGW,MGLi1,MGLi2}.
Hence, we are very curious about the possible impact of the SL test on cosmological constraints for MG theories.
If the SL test can also effectively improve the constraint results of MG models, we may further affirm the conclusion in Ref.~\cite{msl3} that the improvement of parameter estimation by SL test data is independent of the cosmological models in the background.
In this paper, we focus on one popular type of MG theory, $f(T)$ theories. In $f(T)$ gravity, the torsion scalar $T$ in the Lagrangian density is replaced by a generalized function $f(T)$, thus producing a possible mechanism for the cosmic acceleration. We take two typical models as examples, the power-law model $f(T)_{PL}$ \cite{FT1} and the exponential-form model $f(T)_{EXP}$ \cite{FT2}, and quantify the impact of redshift drift measurement on parameter estimation for these models.

We adopt the $f(T)_{PL}$ model with the form
\begin{equation} \label{eq:fT1}
f(T) = \alpha (-T)^{n}.
\end{equation}
Here $n$ is the model parameter, and % $\alpha$ can be fixed as
\begin{equation}
\alpha = (6 H_{0}^{2})^{1-n} \frac{1-\Omega_{m}-\Omega_{r}}{2n-1},
\end{equation}
where $H_0$ is the Hubble constant. $\Omega_m$ and $\Omega_r$ are the present-day density parameters for the matter and radiation, respectively.

We adopt the $f(T)_{EXP}$ model with the form
\begin{equation} \label{fT2}
f(T) = m T_0 \big(1-e^{-p\sqrt{T/T_{0}}}\big).
\end{equation}
Here $p$ is the model parameter, $T_0 = -6H^{2}_{0}$, and
\begin{equation} \label{c}
m=\frac{1-\Omega_{m}-\Omega_{r}}{1-(1+p)e^{-p}}.
\end{equation}

\section{Methodology}

We first constrain the $f(T)$ models using a combination of current data and then choose the best-fit models as fiducial models in producing 30 mock SL test data.
Finally, we constrain the $f(T)$ models again using the simulated SL test data combined with the current data and quantify the improvement in the parameter estimation.

In our analysis, the most typical and commonly used current observations are chosen, i.e., the type Ia supernovae (SNe), the cosmic microwave background (CMB),
the baryon acoustic oscillation (BAO), and direct measurement of the Hubble constant $H_0$.
We use the SNLS3 compilation with a sample of 472 SNe~\cite{snls3} for the SN data.
We use the BAO data presented in Ref.~\cite{wmap9}, i.e., the $r_s/D_V(z)$ measurements from the
6dFGS~\cite{6dF}, SDSS-DR7~\cite{DR7}, SDSS-DR9~\cite{DR9}, and
WiggleZ~\cite{WiggleZ} surveys.
Because we focus on the geometric measurements in this work, for the CMB data, we use the Planck distance priors in Ref.~\cite{WW}.
 We use the direct measurement of $H_0$ from the Hubble Space Telescope,
$H_0=73.8 \pm 2.4$ km s$^{-1}$ Mpc$^{-1}$~\citep{Riess2011}.

In the redshift drift measurement, the redshift variation is defined as the spectroscopic velocity shift~\cite{loeb}
\begin{equation}\label{eq1}
\ \Delta v \equiv \frac{\Delta z}{1+z}=H_0\Delta t_o\bigg[1-\frac{E(z)}{1+z}\bigg],
\end{equation}
where $\Delta t_o$ is the time interval of the observation. 
$E(z)=H(z)/H_0$ is determined by specific $f(T)$ models; for details, see Ref.~\cite{MGW}.

According to a Monte Carlo simulation in Ref.~\cite{Liske}, the uncertainty of $\Delta v$ can be expressed as
\begin{equation}\label{eq2}
\sigma_{\Delta v}=1.35
\bigg(\frac{S/N}{2370}\bigg)^{-1}\bigg(\frac{N_{\mathrm{QSO}}}{30}\bigg)^{-1/2}\bigg(
\frac{1+z_{\mathrm{QSO}}}{5}\bigg)^{x}~\mathrm{cm}~\mathrm{s}^{-1},
\end{equation}
where $S/N$ is the signal-to-noise ratio defined per 0.0125 $\mathring{\rm A}$ pixel, and the last exponent is $x=-1.7$ for $2<z<4$ and $x=-0.9$ for $z>4$. $N_{\mathrm{QSO}}$ is the number of observed QSOs, and $z_{\mathrm{QSO}}$ denotes their redshift. We simulate $N_{\mathrm{QSO}}=30$ SL test data uniformly distributed over six redshift bins of $z_{\mathrm{QSO}}\in [2,~5]$.
We calculate the central values of the mock data by substituting the obtained best-fit parameters in the fit to the current data into Eq.~(\ref{eq1}). The error bars can be computed from Eq.~(\ref{eq2}).

\section{Results and discussion}

\begin{table*}\tiny
\caption{ Fit results for  the $f(T)_{PL}$ and $f(T)_{EXP}$  models using the
current only and current+SL 30-year data. We quote $\pm 1\sigma$ errors, but
for the parameters that cannot be well constrained, we quote the 95.4\% CL lower limits.}
\label{table1}
\small
\renewcommand{\arraystretch}{1.5}
%\centering
\begin{tabular}{ccccccccccccccccccc}
\\
\hline\hline &\multicolumn{2}{c}{current only} &&\multicolumn{2}{c}{current + SL 30-year} \\
           \cline{2-3}\cline{5-6}
Parameter & $f(T)_{PL}$ & $f(T)_{EXP}$ && $f(T)_{PL}$ & $f(T)_{EXP}$\\ \hline

$\Omega_bh^2$        & $0.0221^{+0.0003}_{-0.0003}$
                   & $0.0224^{+0.0002}_{-0.0002}$ &
                   & $0.0222^{+0.0003}_{-0.0003}$
                   & $0.0224^{+0.0002}_{-0.0002}$
                   \\

$\Omega_ch^2$        & $0.1202^{+0.0026}_{-0.0021}$
                   & $0.1173^{+0.0015}_{-0.0015}$ &
                   & $0.1199^{+0.0021}_{-0.0021}$
                   & $0.1172^{+0.0005}_{-0.0006}$
                   \\

$n$                  & $-0.1967^{+0.1191}_{-0.1782}$
                   & $-$ &
                   & $-0.1880^{+0.0982}_{-0.1171}$
                   & $-$
                   \\

$p$                  & $-$
                   & $>4.0815$ &
                   & $-$
                   & $>4.0483$
                   \\

$\Omega_{m}$         & $0.2871^{+0.0096}_{-0.0095}$
                   & $0.2949^{+0.0088}_{-0.0084}$ &
                   & $0.2866^{+0.0017}_{-0.0022}$
                   & $0.2941^{+0.0020}_{-0.0019}$
                   \\

$H_0$                & $70.40^{+1.25}_{-1.14}$
                   & $68.83^{+0.69}_{-0.68}$ &
                   & $70.41^{+0.51}_{-0.41}$
                   & $68.89^{+0.21}_{-0.22}$
                   \\
\hline
\end{tabular}
\end{table*}

%$\Omega_{m}$ 4.70\%, 4.13\%, 0.97\%, 0.94\%;
%
%$H_0$ 2.40\%, 1.41\%, 0.93\%, 0.44\%;
%
%$n$   108.97\%, 81.29\%

\begin{figure}
\begin{center}
\includegraphics[width=9cm]{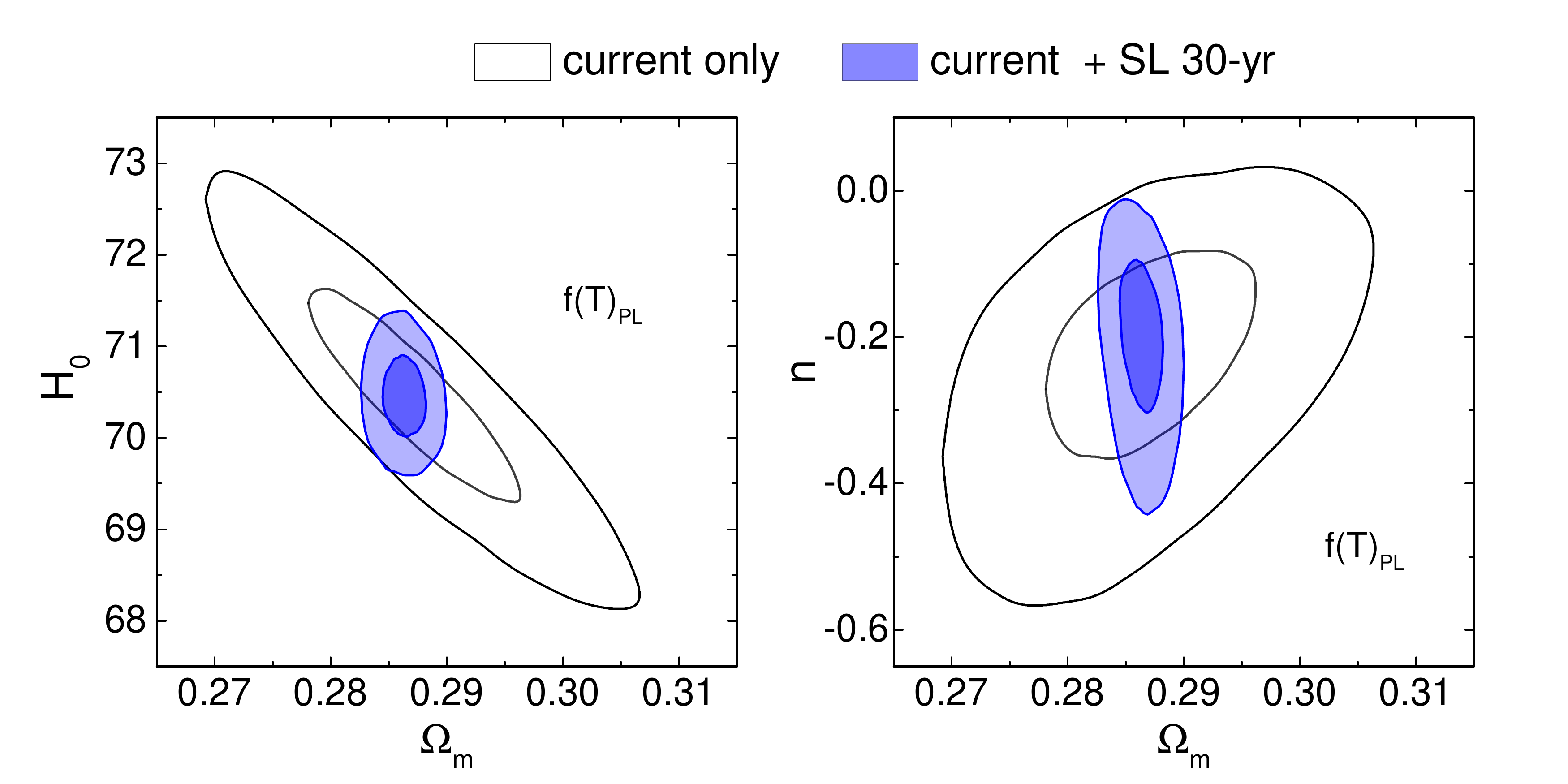}
\end{center}
\caption{Constraints (68.3\% and 95.4\% CL) in the $\Omega_m$--$H_0$ plane and in the $\Omega_m$--$n$ plane for $f(T)_{PL}$ model with current only and current+SL 30-year data.}
\label{fig1}
\end{figure}

Table~\ref{table1} gives the detailed fit results for the $f(T)_{PL}$ and $f(T)_{EXP}$  models using only the current data (current-only) and the current data plus the 30-year SL data (current+SL 30-year).
 We quote $\pm 1\sigma$ errors, but
as the parameter $p$ cannot be well constrained, we quote the 95.4\% confidence level (CL) lower limits.
The joint constraints on the $f(T)_{PL}$ model in the $\Omega_m$--$H_0$ and $\Omega_m$--$n$ planes are shown in Figure~\ref{fig1}.
The 68.3\% and 95.4\% CL posterior distribution contours are shown, where the current-only and the current+SL 30-year results are shown in white and blue, respectively.
For the current-only data, the precisions of $\Omega_m$, $H_0$, and $n$ are constrained to the 4.70\%, 2.40\%, and 108.97\% level, respectively,
whereas for the current+SL 30-year data, the precisions of $\Omega_m$, $H_0$, and $n$ are constrained to the 0.97\%, 0.93\%, and 81.29\% level, respectively.
Using the SL 30-year combined data can clearly improve the precisions of $\Omega_m$ and $H_0$ significantly 
but can improve the constraint precision of parameter $n$ only moderately.

\begin{figure}
\begin{center}
\includegraphics[width=9cm]{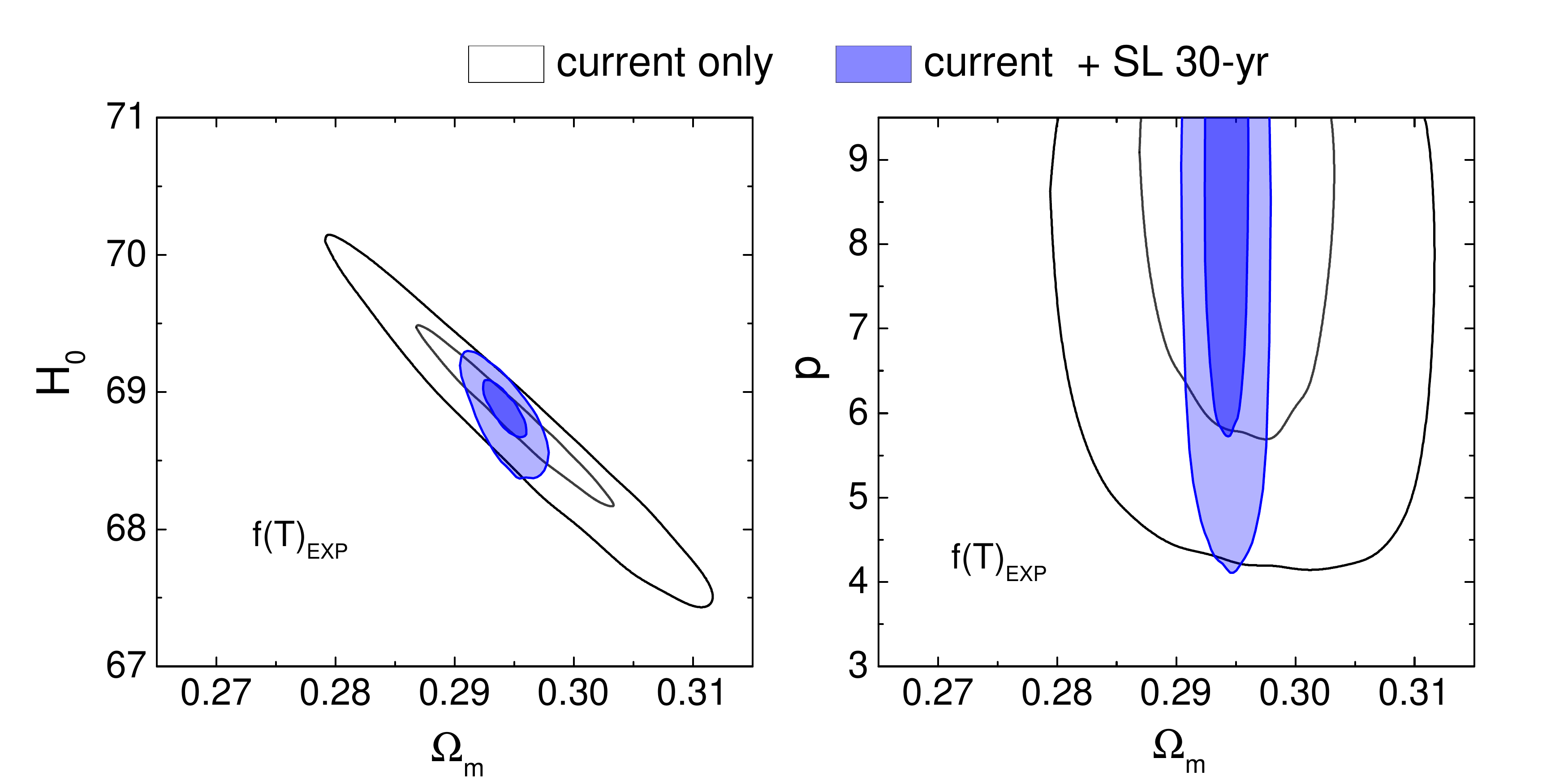}
\end{center}
\caption{Constraints (68.3\% and 95.4\% CL) in the $\Omega_m$--$H_0$ plane and in the $\Omega_m$--$p$ plane for $f(T)_{EXP}$ model with current only and current+SL 30-year data.}
\label{fig2}
\end{figure}

In Figure~\ref{fig2}, we present the joint constraints on the $f(T)_{EXP}$ model (68.3\% and 95.4\% CL) in the $\Omega_m$--$H_0$ and $\Omega_m$--$p$ planes.
The current only and the current+SL 30-year results are shown in white and blue, respectively.
For the current-only data, the precisions of $\Omega_m$ and $H_0$ are constrained to the 4.13\% and 1.41\% level, respectively,
whereas for the current+SL 30-year data, the precisions of $\Omega_m$ and $H_0$ are constrained to the 0.94\% and 0.44\% level, respectively.
The improvements are also remarkable. 
However, these results show that the parameter $p$ cannot be well constrained with either the current data or the current+SL 30-year data,
and that the SL test data cannot affect the fit results of the model parameter $p$.

From the left-hand panels in Figures~\ref{fig1} and~\ref{fig2}, we can conclude that for the considered $f(T)$ models,
future SL test data can efficiently break the strong degeneracy between $\Omega_m$ and $H_0$ existing in the current observational data
and thus can greatly improve the precisions of these parameters.
The results are consistent with those of our previous studies on dark energy models \cite{msl1, msl2, msl3}.
Thus, we can further confirm that the improvement of parameter estimation by SL test data should be independent of the cosmological models in the background.
It is very significant and necessary to include SL test data in future cosmological constraints.

\section{Summary}

The SL test directly measures the temporal variation of the redshift of QSO Lyman-$\alpha$ absorption lines in the so-called ``redshift desert''
($2\lesssim z\lesssim 5$), which is not covered by any other cosmological observation.
In our previous works~\cite{msl1,msl2,msl3}, we performed a serious synthetic exploration of the impact of future SL test data on dark energy constraints. % based on the observation of 30 QSOs.
It was shown that the SL test can break the parameter degeneracies in existing dark energy probes and significantly improve the precision of dark energy constraints.
%Especially, $\Omega_m$ and $H_0$ can be constrained to a high precision for all considered dark energy models with the a 30-year observation of SL test.
In particular, combination with the SL test can constrain $\Omega_m$ and $H_0$ to a high precision for all the considered dark energy models.

In this paper, we quantified the impact of future SL test data on one popular type of MG theory, the $f(T)$ gravity theories.
Taking the power-law model $f(T)_{PL}$ and the exponential-form model $f(T)_{EXP}$ as examples, we found that by using a 30-year observation of the SL test,
the strong parameter degeneracies of $\Omega_m$ and $H_0$ can be effectively broken.
For the $f(T)_{PL}$ model, the precisions of $\Omega_m$ and $H_0$ based only on current data are constrained to the 4.70\% and 2.40\% level,
whereas those based on current+SL 30-year data are constrained to the 0.97\% and 0.93\% level, respectively.
For the $f(T)_{EXP}$ model, the precisions of $\Omega_m$ and $H_0$ based only on current data are constrained to the 4.13\% and 1.41\% level,
whereas those based on current+SL 30-year data are constrained to the 0.94\% and 0.44\% level, respectively.
Thus, the constraint precisions of $\Omega_m$ and $H_0$ can be improved enormously for the considered $f(T)$ models.
However, a 30-year observation of the SL test can improve the precision of the model parameter $n$ for the $f(T)_{PL}$ model only moderately and evidently cannot affect the precision of the model parameter $p$ for the $f(T)_{EXP}$ model.

%In, a 30-year observation of SL test can improve the precision of $\Omega_m$ from 4.70\% to 0.97\%, and that of $H_0$ from 2.40\% to 0.93\%,
%but can only moderately improve the.
%In the, a 30-year observation of SL test can improve the precision of $\Omega_m$ from 4.13\% to 0.94\%, and that of $H_0$ from 1.41\% to 0.44\%,
%but cannot influence on .

The results for $f(T)$ theories are consistent with those for the dark energy models.
We conclude that the improvement of the constraint precision by SL test data is independent of the cosmological models in the background.
To make this conclusion more convincing,
more MG models other than the $f(T)$ models should be explored, such as the Dvali-Gabadadze-Porrati model~\cite{DGP} and different $f(R)$ models~\cite{FR1,FR2}.
We leave a complete analysis of MG theories as future work.

\begin{acknowledgments}
%We acknowledge the use of {\tt CosmoMC}.
This work was supported by the National Natural Science Foundation of
China under Grants No.~11175042 and No.~11522540, the Provincial Department of Education of
Liaoning under Grant No.~L2012087, and the Fundamental Research Funds for the
Central Universities under Grants No.~N140505002, No.~N140506002, and No.~N140504007.

\end{acknowledgments}

\end{document}